# Patent Overlay Mapping: Visualizing Technological Distance

**Luciano Kay**, Center for Nanotechnology in Society, University of California Santa Barbara, Santa Barbara, CA, USA, e-mail: luciano@cns.ucsb.edu

**Nils Newman**, Intelligent Information Services Corporation, Atlanta, GA, USA, e-mail: newman@iisco.com

**Jan Youtie**, Enterprise Innovation Institute & School of Public Policy, Atlanta, GA, USA, e-mail: jan.youtie@innovate.gatech.edu

**Alan L. Porter**, School of Public Policy, Georgia Institute of Technology, Atlanta GA & Search Technologies, Norcross, GA, USA, e-mail: alan.porter@isye.gatech.edu

**Ismael Rafols**,Ingenio (CSIC-UPV), UniversitatPolitècnica de València, València, Spain & SPRU – Science and Technology Policy Research, University of Sussex, Brighton, England, e-mail: i.rafols@ingenio.upv.es

**Abstract**

This paper presents a new global patent map that represents all technological categories, and a method to locate patent data of individual organizations and technological fields on the global map. This overlay map technique may support competitive intelligence and policy decision-making. The global patent map is based on similarities in citing-to-cited relationships between categories of theInternational Patent Classification (IPC) of European Patent Office (EPO) patents from 2000 to 2006. This patent dataset, extracted from the PATSTAT database, includes 760,000 patent records in 466 IPC-based categories. We compare the global patent maps derived from this categorization to related efforts of other global patent maps. The paper overlays nanotechnology-related patenting activities of two companies and two different nanotechnology subfields on the global patent map. The exercise shows the potential of patent overlay maps to visualize technological areas and potentially support decision-making. Furthermore, this study shows that IPC categories that are similar to one another based on citing-to-cited patterns (and thus are close in the global patent map) are not necessarily in the same hierarchical IPC branch, thus revealing new relationships between technologies that are classified as pertaining to different (and sometimes distant) subject areas in the IPC scheme.

**Introduction**

The visualization of knowledge or technological landscapes has been a prominent part of publication and patent analyses since their origins (Hinze, Reiss, & Schmoch, 1997; Small, 1973). However, only in the last decades, improvements in computational power and algorithms have allowed the creation of large maps covering a full database, the so-called global maps of science (see overviews by Klavans & Boyack, 2009; Rafols, Porter, & Leydesdorff, 2010).[i] These science maps or *scientograms* are the visualization of the relations among areas of science using network analysis algorithms.

Visualization procedures for science maps have generally been used to explore and visually identify scientific frontiers, grasp the extent and evolution of scientific domains, and analyze the frontiers of scientific research change (Van den Besselaar & Leydesdorff, 1996). Science mapping efforts have been also used to inspire cross-disciplinary discussion to find ways to communicate scientific progress (see, for example, Mapping Science at http://www.scimaps.org/). Although science maps cannot replace other methodological approaches to data analysis, "visual thinking" can help to interpret and find meaning in complex data by transforming abstract and intangible datasets into something visible and concrete (Chen, 2003). Diverse approaches can be used to create visualizations.

The purpose of this paper is twofold: first, to present the results of a global patent map and, second, to introduce the 'overlay map' technique to locate the relative technological position of an organization's patent activity to support competitive intelligence and policy decision-making. This research draws on the concept of technological distance to interpret linkages among technologies and elaborate a method for a meaningful visualization of technological landscapes.

This visualization approach is a logical extension of the experience with science overlay maps. It draws closely on our previous work on science mapping(Rafols et al., 2010) and opens up new avenues for understanding patent landscapes, which as we will see markedly differ from scientific landscapes. The need for development of tools to benchmark and capture temporal change of organizational innovation activities, or patterns of technological change, also motivates this work. More generally, this new approach also accompanies the broader change from hierarchical, structured knowledge in science and technology (i.e. with subdisciplines and specialties that match departmental structures) to a web of "ways of knowing" resulting from changing social contracts (Gibbons et al., 1994), increasing institutional hybridity (Etzkowitz & Leydesdorff, 2000), and dissonance between epistemic and social structures. Our paper will show that in many instances, technological similarity based on citing-to-cited references is not the same as the hierarchical structures used to organize patented knowledge.

To exemplify the kind of analytical support offered by this approach, this paper illustrates the application of patent overlay maps to benchmark the nanotechnology-related patenting activities of two companies and to reveal the core structure of patenting activities in two different nanotechnology subfields. Nanotechnology is an umbrella term referring to a diverse set of emerging technologies that improve or enable materials, devices and systems using novel properties resulting from the engineering and assembly of matter at extremely small scales. At the nanoscale, scientific discoveries have unveiled novel properties that offer the potential for applications in a wide array of market segments such as energy, pharmaceuticals, and semiconductors. With a wide range of potential applications, nanotechnology is anticipated to have significant business and economic impacts in future years. Our previous work illustrated how science overlay maps help to provide a better understanding of the characteristics and evolution of the nanotechnology field and its subfields (see, for example, Porter & Youtie, 2009; Rafols & Meyer, 2010).

This paper is organized as follows. Section 2 reviews and discusses the concept of technological distance and the analysis of patent literature. Section 3 presents the methodological approach. Section 4 presents preliminary outputs based on the application of patent overlay maps to general patent datasets and the analysis of company patent portfolios and technological fields. Section 5 discusses the advantages and drawbacks of the method and elaborates on next steps and future of patent mapping. The paper also includes information to access supplementary material made available by the authors online as detailed in the Appendix.

**Technological distance and its operationalization**

Technological distance, or the extent to which a set of patents reflects different types of technologies, is a key characteristic in being able to visualize innovative opportunities (Breschi, Lissoni, & Malerba, 2003). Patent documents that reference other patents in similar technology areas have been suggested to offer incremental opportunities to advance an area whereas patent documents that refer across diverse categories may offer the potential for radical innovation (Olsson, 2004). Technological distance is often proxied by patent categories, with patents in a given patent category being considered more similar to one another than to those in other patent categories (Jaffe, 1986; Kauffman, Lobo, & Macready, 2000). For example, Franz (2009) uses patent citations between U.S. patent categories and assigns weights to a patent citing another patent in a different category to reflect a larger technological distance. Hinze et al. (1997) look at co-assignment of multiple IPC categories as a measure of the distance between 30 technological fields. A challenge in relying on patent classifications is that, as technology changes, technology-oriented applications may draw from patents in different hierarchical categories, and subsequently lead to further diversity in patents that cite patents in these categories.Hinze et al's (1997) contribution was important because it established that the global map of patents is similar fordifferent countries (U.S., Japan, Germany) and for different time periods (1982-1985, 1986-1989 and 1990-1993). Given such stability, one can then think of this stable structure as a basemap over which to compare the technological distribution of specific organisations, in the same way that we may compare the distribution of different plant species or multinationals over the world map.

This investigation draws on the concept of technological distance and proposes an alternative approach to relying on administrative patent categories, using patent mapping techniques to visualize technological landscapes based on similarity as indicated through citing-to-cited relationships. A patent map is a symbolic representation of technological fields that are associated with relevant themes. Technological fields are positioned in the map so that similar fields are situated nearby and dissimilar components are situated at a distance. The map is constructed from a similarity matrix based on citing-to-cited patents (i.e. a matrix that reflects similarities amongst IPC categories in how patent cite each other). The similarity measures are calculated from correlation functions among fields according to citations among patent categories. This multidimensional matrix is projected onto a two-dimensional space. Visual output provides for flexibility in interpreting the multidimensional relationships among the patent categories. In addition, this approach allows the user to "overlay" subsets of patent data–representing different types of technological fields, institutions, or geographical regions–to understand the particular technological thrusts and areas of concentration of these entities(Rafols et al., 2010).

Recently other scholars have pursued a similar patent record-level approach to create global maps of technology that characterizes the proximity and dependency of technological areas (see, for example, preliminary work in Boyack & Klavans (2008), and related approaches by Schoen et al. (2012), or Leydesdorff, Kushnir& Rafols (In Press).)[ii] Those efforts have also sought to use the maps to benchmark industrial corporations to inform corporate and policy decision-making. The differences with the approach presented in this paper are primarily related to the definition of categories (which yields different number and composition of technology groups) and the relationships among them (generally based on citation-based co-occurrence of IPC categories, which yields maps with different structures). The Boyack&Klavans (2008) work is based on class (3-digit) level IPC categories, while Leydesdorff, Kushnir& Rafols(In Press) include Class (3-digit) and Sub-Class (4-digit) analyses based on USPTO data rather than EPO. These IPC-based approaches work with the existing classification system, which is a product of patent office history, regardless of the intensive quantity of patents in certain categories. For example, categories such as A61 ("Medical or Veterinary Science") has a very large quantity of patents, while categories such as A42B ("Hats") have very few. This uneven distribution of patents limits visualization ability if using the native classification system as is. The contribution of this work is the development of a patent mapping approach based on IPC categories that corrects this uneven patent distribution as explained below. Schoen et al. (2012) patent map is based in technology-based categorization that combines different IPC branches. As it was the case in science maps (Klavans&Boyack, 2009; Rafols & Leydesdorff, 2009), it is very important to compare the results of diverse global patent maps using different classification and visualization algorithms to test the robustness of patterns observed. Without significant consensus on the shape and relative position of categories, global maps are meaningless as stable landscapes needed to compare organizational or technological subsets.

The approach used in this paper draws on learning from the authors' prior work on science mapping, particularly the trade-off between sufficient detail and not too much detail to be easily visualized by the user. The challenges faced when developing this kind of patent map include gathering patent data in appropriate quantity to create meaningful maps and the choice of an equivalent to citation patterns (because citations may not be functionally equivalent to journal citations) and an equivalent to Web of Science Categories (previously known as ISI Subject Categories,) for which IPC categories may not be suitable analogs. Using IPC categories from patent documents also involves specific challenges, such as deciding on the appropriate level of analysis to obtain satisfactory results. This latter point is related to the IPC classification scheme that offers Sections, Classes, Sub-Classes and Groups from which to choose. While the Sub-Class (i.e., 4-digit IPC) level seems appropriate because of the degree of detail in subject matter definitions, it suffers a "population" problem related with the significant variation of the number of patents classified in each IPC Sub-Class, which is likely to lead to underrepresented technologies in maps. Some Sub-Classes have several hundred thousand patents, whereas others have only a few hundred. Thus, a more appropriate grouping of IPC categories is needed to more evenly represent the number of patents across the patent system.

**Implementation**

This global patent map is based on citing-to-cited relationships among IPCs of European Patent Office (EPO) patents from 2000-2006. This period was chosen because of its stability with respect to IPC 7 categories. IPC 7, at the time we conducted this study, represented the longest period of stable classification, as IPC 8 was just rolling out at the time of this research and could potentially add and/or modify categories. Future work would involve comparing patent overlay maps based on IPC 7 and IPC 8, but first, the project team needed to make sure it could produce mapping process with a stable set of categories. The dataset containing IPCs relationships, extracted from the Fall, 2010, PATSTAT database version, represents more than 760,000 patent records in more than 400 IPC categories. This data range begins with patent EP0968708 (which was published in January 2000) and ends with patent EP1737233 (published in December 2006.) An analysis with this kind of coverage benefits from a relative stability of Version 7 of the patent classification system maintained during the 2000-to-2006 time period.

In this approach, the process of data gathering and pre-processing involves, first, going through each patent record to collect all the instances of IPC categories in the dataset and, second, solving the aforementioned "population" problem. The proposed solution for patent categories with relatively few patents is to fold the IPC category up into the next highest level of aggregation to create relatively similar sized categories. This solution comprises three rules: 1) for IPC categories with large population, use the smallest Sub Group level; 2) for small population IPC categories, aggregate up to General Group level, Sub-Class or Class; and, 3) establish a floor cut-off and drop very small aggregated populations. As a result, IPC categories with instance counts greater than 1,000 in the dataset were kept in their original state. Those categories with instance counts less than 1,000 were folded up to the next highest level until the count exceeded 1,000 or the Class level was reached. During the folding, any other IPC categories with counts exceeding 1,000 in the same branch were left out of the folding count. If at the Class level (i.e. 3-digit), the population was less than 1,000, the IPC code was dropped for being too small to map. Table 1 illustrates this approach for the 4-digit IPC class A61K.

**Table 1. Data pre-processing to group IPC categories, selected examples[1]**

| Original IPC in dataset | Catchwords | Original Record Count |
|---|---|---|
| A61B | Diagnosis; Surgery; Identification | 25,808 |
| Authors' process splits this out into: | | |
| A61B 5/00 | Measuring for diagnostic purposes | 1,415 |
| A61B 17/00 | Surgical instruments, devices or methods, e.g. tourniquets | 1,493 |
| A61B 19/00 | Instruments, implements or accessories for surgery or diagnosis not covered by any of the groups | 1,444 |
| and a remainder: | | |
| A61B[2] | | 21,456 |

Notes: 1. Each IPC with an instance count greater than 1,000 was kept in its original state.
2. Each IPC with an instance count less than 1,000 was folded up to the next highest level until the count exceeded 1,000 or the class level was reached.

This pre-processing (in which the roll-up heuristics were performed through a compiled code written in C++) yields IPC categories at the Class, Sub-Class, Main Group and Sub-Group levels, with levels that ensure broadly similar numbers (i.e. within two orders of magnitude) of patents across categories. Although we keep referring to these categories as 'IPC categories', they are not the standard IPC categories since they have a mixed hierarchical composition. The smallest categories in the dataset have 1,000 patents, with this bottom threshold chosen to yield a sufficient count for statistical analyses. The largest category—A61K (defined as "Preparations for Medical, Dental, or Toilet Purposes") but subtracting 16 seven-digit IPCs with more than 1,000 patents each—has more than 85,000 patents. The initial implementation actually involved testing several cut-off values (e.g. 700, 1,000 and 1,500 records) that yielded different numbers of IPC categories. The cut-off at 1,000 was deemed suitable for this analysis, as it seems to provide a sensible compromise between accuracy of the fields, and readability in the map. This choice produces 466 IPC categories that are mapped to a thesaurus for data pre-processing. Out of these categories, 44 categories (representing 2.78 million patents) remain at the Class (3-digit) level, 297 categories (representing 29.11 million patents) remain at the Sub-Class (4-digit) level, 56 at categories at the Main Group level (representing 5.10 million patents) and 69 at the Sub-Group level (representing 4.75 million patents.)

**Table 2. Number of categories and patents obtained with the multi-level aggregation process**

| Level in Classification[1] | # Categories | Mean # apps[2] | % of apps[3] |
|---|---|---|---|
| Class (3 digit) | 44 | 63,280 | 6.7% |
| Sub-Class (4 digit) | 297 | 97,997 | 69.7% |
| Main Group (7 digit, \00) | 56 | 91,144 | 12.2% |
| Sub-Group (7 digit, \##) | 69 | 68,781 | 11.4% |
| Total | 466 | 89,569 | 100.0% |

Notes: 1. See www.wipo.int for more information about these levels; 2. Mean number of patent applications. 3. Share of patent applications in the dataset.

The next step involves extracting from PATSTAT the patents cited by the target records. The IPCs of those patents are mapped to the 466 IPC categories. Some of the patents cited by those in our IPC 7 dataset were published under previous categorization systems; however, this spillover does not lead to any problems from a categorization standpoint because IPC integrates prior categorizations into more recent versions. The result of this data collection allows the creation of a table containing, in each row, sets of Patent Number, IPC Number, Cited Patent Number, andCited IPC Number. This data table has been further processed and saved in an appropriate file format for the next step using the software *Pajek*. This software also helped to create the global map and individual overlay maps for examples of companies and technological fields.

The final data processing steps involve generating a cosine similarity matrix among citing IPC categories (using conventional cosine similaritynormalized by the square root of the squared sum,) and then factor analysis of the IPC categories (following the method used in global science maps by Leydesdorff and Rafols (2009). A factor analysis of the citing-to-cited matrix among IPC categories is then used to consolidate the 466 categories into 35 "macro patent categories." No distinction was made between primary and second classifications and all citing-to-cited relationships were counted equally (i.e. without fractional counting.)We tested different factor solutions from 10 to 40. The 35-factor solution appeared to provide a sensible and convenient classification of the IPC categories. These 35 factors form the basis for color-coding the 466 categories that are represented in visualizations. The list of 35 factors is available in Supplementary File 1 (see details at the Appendix).The visualizations also require converting IPC codes to succinct text labels, which we did by shortening lengthy IPC definitions. Therefore, labels may not fully capture all the technologies within a category. These IPC category labels were then used as a basis for creating descriptors for each factor as shown in the maps (next section.)

The creation of patent overlay maps using a wide range of IPC-based categories requires consideration of the classification system of reference. This research draws on the IPC 7 classification system that, compared to previous versions, includes class codes such as B82B that are relevant to the nanotechnology domain. The IPC 7 system is also more stable than the more recent IPC 8, but still received some updates during the time-period relevant to this study, including the addition of the B82B technology classification. Those updates do not affect the structure of the maps because the newly added classifications represent a small number of patents (i.e. below our cut-off value) and do not affect the map-based analyses because patent records in newly added classifications are generally assigned to other technology categories as well.[iii] Future developments of these maps will require updating the thesaurus developed to match the 466 categories of the global patent maps.

**Test and preliminary results**

The global patent map

The full map of patents shows all 466 categories in a Kamada-Kawai layout (using Pajek) that represents technological distances and groups of technologies in each of the **35factors or technological areas**shown with the same color (Figure 1). Label and color related settings were adjusted to produce a reasonably clear map and facilitate its examination. The map suggests three broad dimensions of patenting interrelationships based on the overall position of technological areas. The left side of the map represents bio-related patents, including food, medicine and biology. The lower right part of the map includes semiconductor, electronics, and information & communications technologies (ICT). The upper right portion of the map is primarily comprised of automotive and metal-mechanic related technology groups.

**Figure 1. Full patent map of 466 technology categories and 35 technological areas**

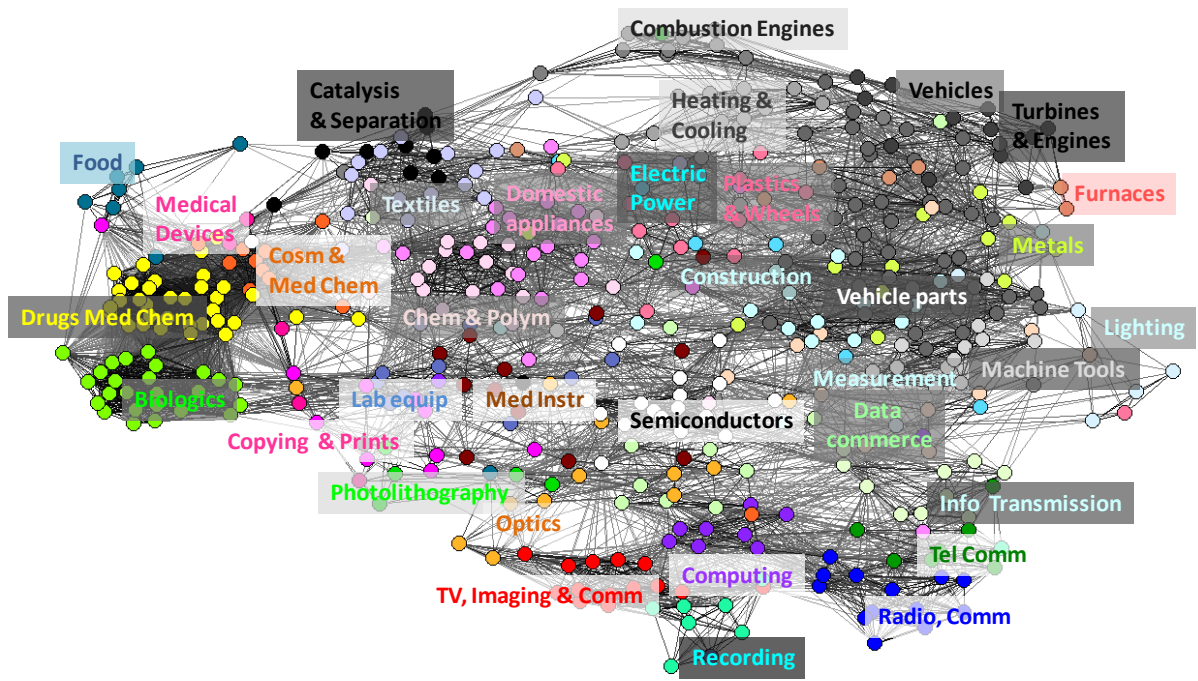

Note: each node color represents a technological area; lines represent relationships between technology categories (the darker the line the shorter the technological distance between categories;) labels for technological areas are placed close to the categories with largest number of patent applications in each

area. Higher resolution figures can be found in supplementary file 2, as indicated in the Appendix. An interactive version of the map is available here:
http://www.vosviewer.com/vosviewer.php?map=http://www.sussex.ac.uk/Users/ir28/patmap/KaySupplementary3.txt

Difference between hierarchy and similarity

A closer look shows that the structure of the map reflects technological relationships across the hierarchical administrative boundaries of the subject matter specifications in the IPC scheme. While counts of IPC sections (i.e. the first letter of IPC codes, A, B, C, D, E, F, G, H) are commonly used as a measure of technological distance in patents, the 35 technological areas that are derived from cross-citations in our patent map often span multiple sections. For instance, the Vehicles area includes six different sections, and the Heating and Cooling, Construction, and Metals areas include five different sections. Textiles, Lighting, Semiconductors, and Chem and Polymers include four different sections. Eleven technological areas (Measurement, Domestic Appliances, Plastics and Wheels, Photolithography, Optics, Copying and Printing, Catalysis and Separation, Lab equipment, Cosmetics and Med Chem, Biologics, Drugs and Med Chem) include three different sections. Ten areas (Turbines and Engines, Machine Tools, Furnace, Electric Power, Info Transmission, Data Commerce, Med Instruments, Combustion Engines, Telephone Comm, TV, Imaging &Comm) have two different sections. Only Medical Devices, Food, Recording, Computing, and Radio Communication areas encompass a single section (further details on this are available in Supplementary File 1.)

This difference between hierarchy and similarity can be observed by comparing Figure 1 with the same map with the nodes colored according to the eight major IPC sections (Figure 2). This observation is strong evidence that the IPC classification on its own is not an appropriate framework to investigate technological diversity without taking account technological distance.

**Figure 2. Full patent map of 466 technology categories and eight 1-digit IPC classes**

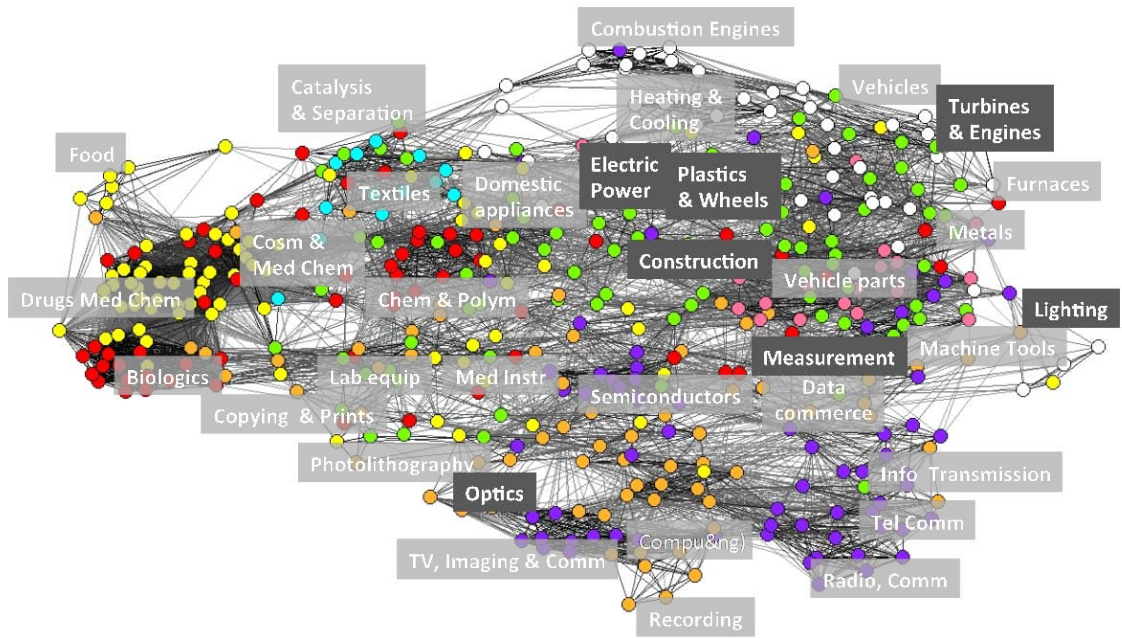

Note: each node color represents a 1-digit IPC section; lines represent relationships between technology categories (the darker the line the shorter the technological distance between categories).Yellow is for Section A: Human Necessities; Light green is for Section B: Performing Operations, Transporting; Red is for Section C: Chemistry, Metallurgy; Light blue is for Section D: Textiles, Paper; Light red is for Section E: Fixed Constructions; White is for Section F: Mechanical Engineering, Lighting, Heating, Weapons, Blasting; Orange is for Section G: Physics; Blue is for Section H: Electricity.

We can further elucidate the classification underlying our map by relating these categories to those in another prominent patent map. The map developed by Leydesdorff and colleagues uses Classes (3-digit) and Sub-Classes (4-digit)(refer to http://www.leydesdorff.net/ipcmaps/).In contrast, our map uses more detailed categorizations to disaggregate some of the patent groupings into more fine-grained analyzable components. By way of example, Leydesdorff'sClass-based (3-digit)map has a single node representing "Medical or Veterinary Science" (IPC A61) on the bottom left side of the map. However, because A61 includes a large share (well over twenty percent) of patents, our map disaggregates this "super-node" into 53 different nodes which end upinfive different medical/veterinary science related clusters or technological areas: (1) Drugs, Med. Chemistry, (2) Biologics, (3) Cosmetics and Med. Chemistry, (4) Medical Instruments, and (5) Medical Devices. Each of these areas is made up of categories that come from different sections and are classified at different levels. Table 3 illustrates how these multiple levels co-exist for the case of the technological area that we labeled "Biologics."This area includes categories at the Class level ("Agriculture", A01,) Sub-Class ("Peptides", C07K,)Main Group ("Peptides, medical", A61K 38/00) and Sub-Group ("Recombinant DNA", C12N 15/09.)

**Table 3. List of categories for the technological area "Biologics"**

| Category Label | IPC number | # Apps |
|---|---|---|
| Agriculture | A01 | 45,126 |
| Animal husbandry | A01K | 14,548 |
| Peptides, medical | A61K38/00 | 482,120 |
| Antigens | A61K39/00 | 20,010 |
| Antibodies | A61K39/395 | 47,662 |
| Gene therapy | A61K48/00 | 15,899 |
| Saccharides | C07H21/00 | 14,578 |
| Peptides, compounds | C07K | 58,219 |
| Peptides from humans | C07K14/435 | 43,462 |
| Peptides from animals | C07K14/47 | 14,602 |
| Immunoglobulins | C07K16/18 | 27,481 |
| Extractions from organisms | C12N | 26,627 |
| Modified fungi | C12N1/15 | 47,884 |
| Modified yeasts | C12N 1/19 | 32,469 |
| Cellulose processes | C12N 1/21 | 13,631 |
| Virus transformed cells | C12N 5/10 | 10,402 |
| Recombinant DNA | C12N15/09 | 21,345 |

| Genes encoding animal proteins | C12N15/12 | 25,010 |
| Fermentation for food | C12P | 29,202 |
| Testing, microorganisms | C12Q | 18,442 |
| Testing, nucleic acids | C12Q1/68 | 22,731 |
| Bacteriology | C12R | 48,984 |
| Measuring biological material | G01N33/50 | 517,367 |
| Immunoassay | G01N33/53 | 43,835 |
| Measuring using proteins, amino acids, lipids | G01N33/68 | 119,957 |

Note: An equivalent table for each of the 35 technological areas can be find at Supplementary File 1, under the tab "Label and Count Table".[iv]

Given our map's ability to present disaggregated categories, we are able to show 35 technological areas or clusters versus only five in the Leydesdorff map. This more disaggregated clustering enables differentiation of the patent portfolios of a company engaged in cosmetics patenting from one engaged in drug development and from yet another engaged in medical instrument development. In conclusion, we believe that the multi-level method of classification proposed here achieves a more accurate description than a straightforward use of IPC classes at the Class or Sub-Class level.

A major problem in these comparisons is that in the case of patents, unlike the map of science, where there has been a pre-established conventional understanding of disciplines, it is not clear how groups of technologies can be interpreted. This problem is compounded by the heterogeneous nature of the patent classes, which includes materials (e.g. "Alloys", C22C,) devices (e.g. "Machines and engines", F01) and products (e.g. "Ships", B63.)This conceptual diversityis observed within the technological classes derived from the factor analysis. For example the area of "Turbines and engines" includes "Turbines" (F01D,) "Jet propulsion" (F02K,) "Aircraft equipment" (B64D) and "Airplanes and helicopters" (B64D)—elements from distinct branches of the IPC classification. These four subclassesobviously co-occur but rather thanbeing similarthey likely co-occur because they areembedded and/or complementary.And the area of "Lightning" includes "Basic electric elements" (H01,)Lighting (F21,) "Vehicle signaling" (B60Q) and "Specialized equipment used in roads" (E01F)—which again are complementary pieces of technology at different levels of aggregation. This difficulty we are facing is not simply a problem of classification, but a conundrum due to the multiple meanings and scales that the technology concept may take (Arthur 2010, p. 28).

Awareness of the conceptual heterogeneity of nodes or elements in the map raises the issue of whether the maps show "similarity" between categories as we have assumed, or other properties such as co-occurrence and complementarity. For example, patents of metals and automobiles are related not because these categories are similar but because automobiles are often made of metals. Also, plastics and metals may co-occur simply because they are materials that are used in similar products such as buckets and automobiles, not because they are similar.[v]This issue suggests that the interpretation of the patent map should be ontologically flexible. In other words, when interpretingthis patent map, one should take into account that both the elements and the relations may have different meanings.

Comparison of map structures

The overall structure of our map appears to be consistent with previous technological maps based on patents that used different algorithms for aggregating IPC categories (Hinze et al., 1997, Boyack & Klavans, 2008; Leydesdorff et al., 2012; Schoen et al., 2012).Hinze's map offers the most straightforward comparison given that it only has 35 categories. The similarity in position to our map with the technological areas in the extreme ends of the network is quite striking. For example, in the bioscience pole, categories related with food, drugs and biotechnology have the same relative position. In the ICT pole as well, optics, audiovisual technologies and telecommunications also keep the relative position. "Semiconductors" however occupies a more central position in our map than in Hinze's, resemblingthe map developed by Boyack but not the one byLeydesdorff. In the vehicle/mechanical pole, one can relate Hinze's categories to our technological areas (engines, turbines, mechanical elements, transport,) but it is not possible to compare the relative positions of the two maps due to the lack of comparable labels in the categories. It is worth stressing that Hinze's visualisation is based in multidimensional scaling, whereas the one presented in Figure 1 is achieved with Kamada-Kawai algorithm; hence, the similarities arise from factors other than the particular layouts used in each case.

Based on these similarities between Hinze's and Figure 1, we are confident that our map captures the main axis in the broad relative position of technologies. The map that differs the most, among those inspected, is the one by Schoen et al. (2012), which nevertheless still partially captures the three axes mentioned. The difference in Schoen's map might reflect a different layout algorithm rather than substantive differences in the relations between technologies. In sum, further research is needed to better understand the relative structure of technologies, and ascertain whether the structure observed is robust and stable (as it was surprisingly found in the map of science, see Klavans and Boyack, 2009), or whether it is susceptible to differences stemming from the use of different (still equally valid) presentation angles.

Interconnectedness

Another interesting feature of the global patent map is the high level of interconnectedness of most of the 35 technological areas. This can be observed not only in many connections among technology groups within each technological area, as shown by the densest areas of the map, but also across them. Some exceptions are areas such as Food, Drugs & Med Chem, Biologics, TV Imaging &Comm, Cosm& Med Chem, and Radio &Comm that form more uniform clusters of technology groups (i.e. they appear as clusters of nodes of the same color) (Figure 1). Another notable feature is the short distance among technologies in a handful of groups such as Drugs & Med Chem and Biologics, as shown by denser areas and darker lines in the left hand side of the maps. The sparse areas of the map are those associated with technological areas that comprise fewer technology categories include, for example, Electric Power, Lighting, and Recording.

Patent overlay maps

Based on the global patent map, patent overlay maps allow, for example, benchmarking of companies and specific technological fields. To illustrate and test the application of patent map overlays, two corporate datasets of nanotechnology patent applications have been created for Samsung and DuPont, and two nanotechnology subfield datasets have been created for Nano-Biosensors and Graphene nanotechnology applications, using data from the Georgia Tech Global Nanotechnology databases in the same time period (2000-2006).

The visual examination of maps shows nanotechnology development foci that vary across companies (even for those in similar industry sectors) and different patenting activity levels for the studied period. The two overlays presented herein appear diversified and encompass a number of technological areas. The patent overlay created for Samsung, for example, shows activity concentrated on semiconductors and optics, with a notable level of patenting activity across other areas as well (Figure 3a). The company has also some prominent activity on technological areas broadly defined as Catalysis & Separation, Photolithography, and Chemistry & Polymers. The focus of DuPont (Figure 3b), on the other hand, is more on Drugs, Medicine & Chemistry, Chemistry & Polymers, and Biologics. This company seems to have a portfolio of patent applications that is even more diversified, but it also is less active in terms of patenting activity, than Samsung.

**Figure 3. Patent overlays applied to company benchmarking**

   a) **Samsung**

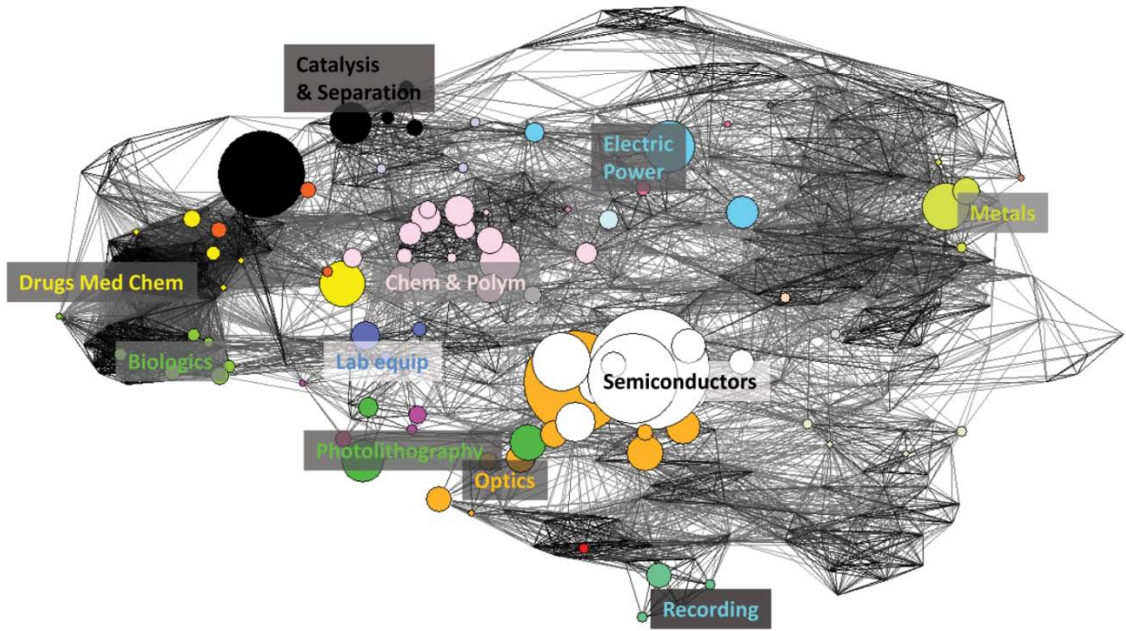

   b) **DuPont**

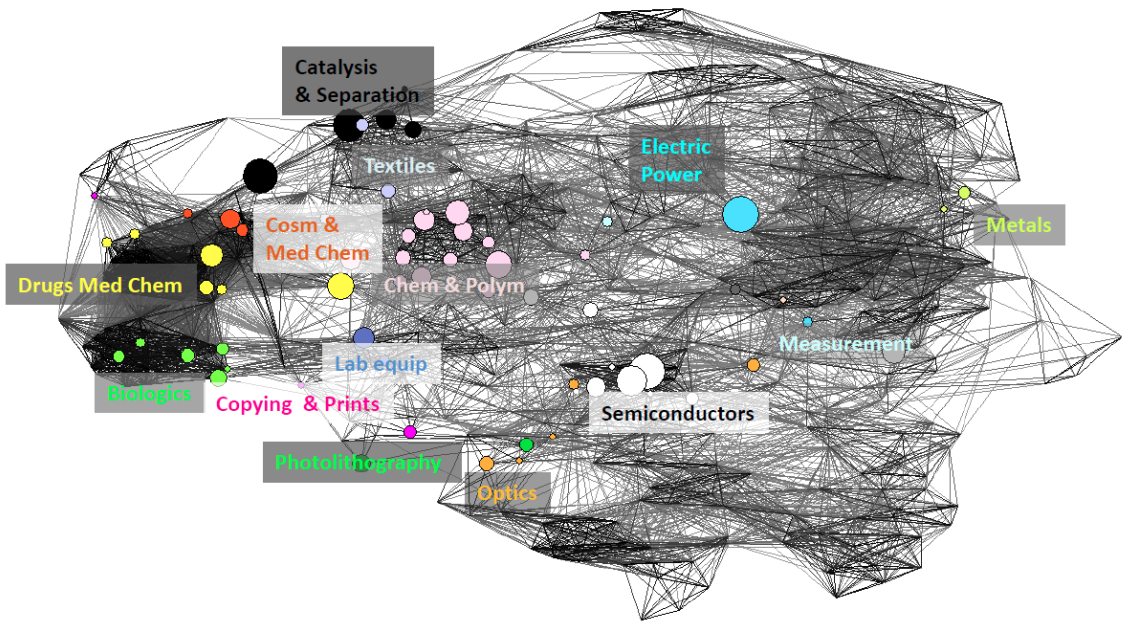

Note: labels shown only for top technological areas of the company patent portfolio; the size of nodes is proportional to the number of patent applications in the corresponding technology group. Higher resolution figures can be found in supplementary file 2, as indicated in the Appendix.

The application of patent overlays to the analysis of technological subfields can also help provide a better understanding of technologies involved in the development of these subfields and relationships between them and with the patent portfolio of companies. Yet, while the patent maps applied to companies reflect the result of a corporate strategy implemented by a single organization, patent maps applied to technological fields reflect the aggregation of activities of multiple (and usually numerous) categories in the same or different sectors.

In the application of patent overlay maps to nanotechnology, technological developments in nano-biosensors are focused on categories such as Laboratory Equipment, Semiconductors and Biologics (Figure 4a). The subfield of Graphene, a more recent development that was recognized with the 2010 Nobel Prize in Physics, presents lower activity levels with a diversified focus on Catalysis & Separation, Chemistry & Polymers, Semiconductors and Optics among others (Figure 4b).

# Figure 4. Patent overlays applied to field mapping

**a) Nano-biosensors**

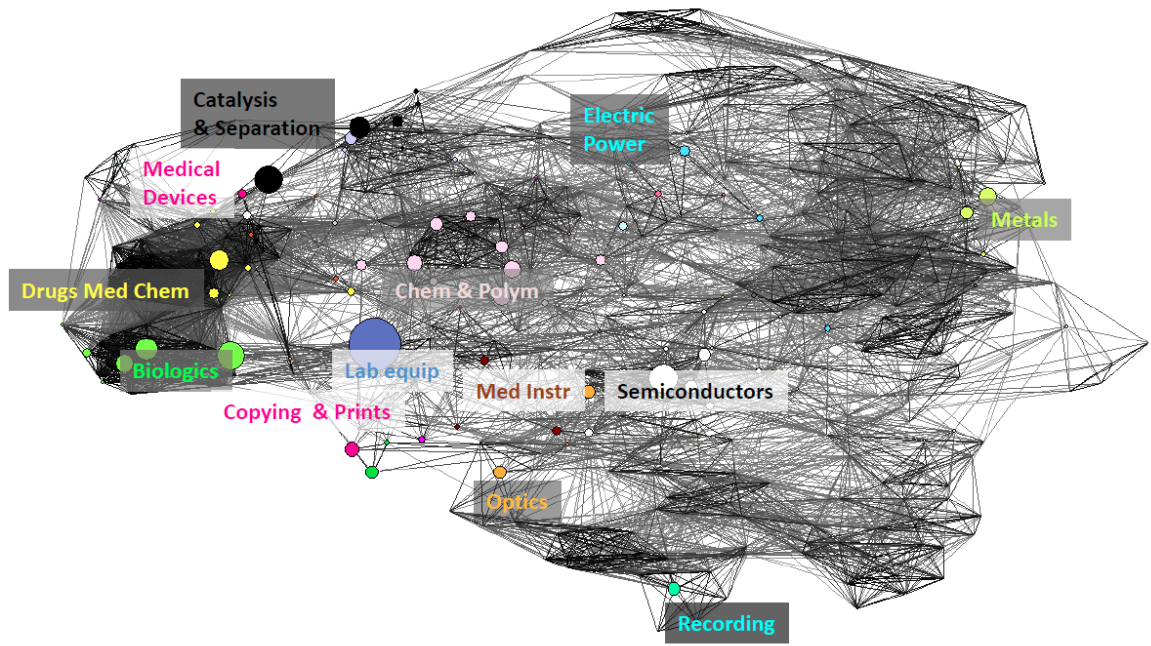

**b) Graphene**

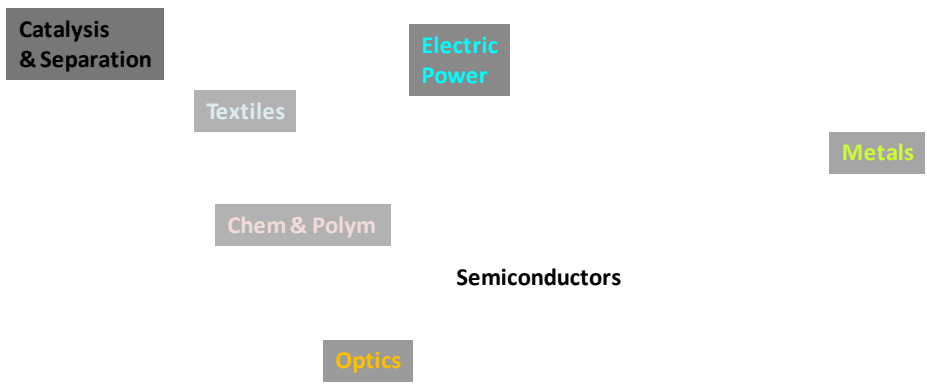

Note: labels shown only for top technological areas in the subfield; the size of nodes is proportional to the number of patent applications in the corresponding technology group. Higher resolution figures can be found in supplementary file 2, as indicated in the Appendix.

**Conclusion**

This paper presents preliminary results of a new patent visualization tool with potential to support competitive intelligence and policy decision-making, following a methodology successfully used in science overlay mapping (Rafols et al., 2010). The approach involves a two-step visualization process. First, we build a global map that shows the technological distance among patent categories using citing-to-cited information for seven years of EPO data. Second, we overlay the patenting activity of specific organisations or in specific technological fields over the fixed "backbone" of the patent map. The aim of this superposition or overlay, is to help understand the patent portfolio of an organisation in the context of the overall technological landscape.

The approach offers distinctive visualization capability with parsimony. In contrast to prior IPC-based global patent maps, this approach recombines IPC categories to reflect a finer distribution of patents. Thus, it enables improved differentiation ability in categories with a large amount of patenting activity such as "Medical or Veterinary Science" (IPC A61).

The definition of categories and its implementation using a thesaurus to match IPC categories facilitates replication by helping to trace back individual categories to verify results and make improvements. Nevertheless, these maps are only reliable to the extent that assignation of patents to IPC categories is accurate and meaningful. Since patent assignation to IPCs may not always be accurate, a large set of patents may be required to ensure that the portfolio of patents shown in an overlay map can be trusted to convey the patenting activities of an organisation represented (in the case of science maps, this was estimated to be above 1,500 publications for high resolution accuracy, and above 100 publications for lower resolution) (see Appendix 1 in Rafols et al. (2010).)

One of the most interesting findings is that IPC categories that are close to one another in the patent map are not necessarily in the same hierarchical IPC branch. This finding reveals new patterns of relationships among technologies that pertain to different (and sometimes distant) subject areas in the IPC classification. The finding suggests that technological distance is not always well proxied by relying on the IPC administrative structure, for example, by assuming that a set of patents represents substantial technological distance because the set references different IPC sections. This paper shows that patents in certain technology areas tend to cite multiple and diverse IPC sections. For example, the Drugs & Medicine and Biologics dimensions include various drug-related Sub-Classes in IPC Class A61, but they also include several chemistry compound Sub-Classes in IPC Class C07; traditional measures would assume that technologies in these dimensions are distant because they include two different sections (sections A and C), but our network map shows that technologies in these two sections are closely interrelated, inasmuch as the patents in these Sub-Classes tend to cite one-another. An improved measure of technological distance would take into consideration patent citation or co-occurrence characteristics.

Potential applications of patent overlay maps include organizational and regional/country benchmarking (e.g. for the examination of competitive positions,) exploration of potential collaborations, and general analysis of technological changes over time. For example, the comparison of maps over time can reveal new patterns of relationship among categories that might help to understand the emergence of new fields and the extent of their impact. Patent maps may also reveal relatively unexplored technological areas that are more central to other technologies or highlight denser areas with more technological interdependency that might form platforms for the emergence of future technology applications (like the Drugs & Medicine and Biologics categories in the maps shown in this paper.) Most of these explorations may require greater granularity for such analysis and policy decision-making (except in the case of large firms with extensive patent portfolios, such as the example of Samsung and DuPont illustrated). This need for granularity is a challenge that faces all global maps. Future work would enable greater ability to drill-down in certain areas, as well as to compare different global maps—for example, maps based on IPC 8 with maps based on IPC7 version—but a stable global map is required as an initial base for such an effort.

Ongoing work has sought to overcome some issues found in the development of the original patent overlay maps. Among the most important issues is the coverage of the thesaurus developed to match 466 IPC categories based on the main patent dataset. While this dataset covers a wide range of IPC categories, the resulting thesaurus still does not match a number of IPC categories in the datasets created for patent overlay maps. This kind of issue varies across patent overlay datasets and may represent a significant proportion of the patent records in certain cases. This is, however, a problem that can be solved in future implementations by creating a new thesaurus based on a larger dataset that covers more than seven years of patent activity.

Next steps in this research thrust include updates of the basemap based on the current version of the PATSTAT database and use of the most recent IPC classification, version 8. Refining the patent database to focus only on patent grants (it currently includes applications as well as grants) is one path for future work, while another is to develop a patent map for patents from other patent authorities besides EPO. In addition, the stability of the patent maps could be tested with the segmentation of maps by year or year ranges. The backbone patent map in this paper should be compared with results from other global patent mapping efforts to determine the extent of consistency between these maps. Although we have presented some preliminary comparisons, a more rigorous and systematic approach for comparing these maps and categorizations is needed (see, for example,Klavans&Boyack, 2009; Rafols & Leydesdorff, 2009). Potential future research includes the analysis of connections between patent maps and science maps, with particular focus on technological fields with strong science links [vi], as well as classifications based on full clustering of an entire database rather than a subset (as recently done in science with more than 10 million records by Waltman& van Eck, 2012).

**Acknowledgments**

We are grateful to Kevin Boyack, Loet Leydesdorff and Antoine Schoen for open and fruitful discussions about this paper. This research was undertaken largely at Georgia Tech drawing on support from the US National Science Foundation (NSF) through the Center for Nanotechnology in Society (Arizona State University; Award No. 0531194); and NSF Award No. 1064146 ("Revealing Innovation Pathways: Hybrid Science Maps for Technology Assessment and Foresight"). Part of this research was also undertaken in collaboration with the Center for Nanotechnology in Society, University of California Santa Barbara (NSF Awards No. 0938099 and No. 0531184).The findings and observations contained in this paper are those of the authors and do not necessarily reflect the views of the US National Science Foundation.

**Appendix: Supplementary materials**
The authors made available three supplementary online files:

- **Supplementary File 1** is an MS Excel file containing the labels of IPC categories, citation and similarity matrices, factor analysis of IPC categories. It can be found at:
  http://www.sussex.ac.uk/Users/ir28/patmap/KaySupplementary1.xls
- **Supplementary File 2** is an MS PowerPoint file with examples of overlay maps of firms and research topics. It can be found at:
  http://www.sussex.ac.uk/Users/ir28/patmap/KaySupplementary2.ppt
- **Supplementary File 3** is an interactive version of map in Figure 1visualized with the freeware VOSviewer. It can be found at: http://www.vosviewer.com/vosviewer.php?map=http://www.sussex.ac.uk/Users/ir28/patmap/KaySupplementary3.txt

[i] Lately, there has been a proliferation of global maps (see, for example, Bollen et al., 2009; Boyack, Börner, & Klavans, 2009; Boyack, Klavans, & Börner, 2005; Janssens, Zhang, Moor, & Glänzel, 2009; Leydesdorff & Rafols, 2009; Moya-Anegon et al., 2004; Moya-Anegón, Vargas-Quesada, Chinchilla-Rodríguez, Corera-Álvarez, & Herrero-Solana, 2007; Rosvall & Bergstrom, 2010).

[ii] Thomson Reuters also has a patent visualization capability, Aureka, but it is a local rather than a global mapping application.

[iii] The analysis shows that only 0.2 percent of the patents of Samsung and 2.6 percent of the patents of Dupont that are solely assigned to the B82B class are not represented in the maps.

[iv] Available at: http://www.sussex.ac.uk/Users/ir28/patmap/KaySupplementary1.xls

[v] We thank Antoine Schoen for this point.

[vi] See for example http://www.mapofscience.com for an overlay of patents in the map of science carried out by Kevin Boyack and Richard Klavans (unpublished). Accessed September 23rd 2013.